\pdfoutput=1
\documentclass{article}

\newif\ifarxiv\arxivtrue

\usepackage{microtype}
\usepackage{graphicx}
\usepackage{subfigure}
\usepackage{booktabs}
\usepackage{array}
\usepackage{xcolor}
\usepackage{amsmath}
\usepackage{multirow}
\usepackage{hyperref}
\usepackage{xurl}
\usepackage{tikz}
\usetikzlibrary{positioning,arrows.meta,decorations.pathreplacing}

\ifarxiv
\usepackage[T1]{fontenc}
\usepackage[accepted]{mlsys2025}
\makeatletter\renewcommand{\mlsys@appearing}{Preprint.}\makeatother
\hypersetup{pdfsubject={Preprint}}
\else
\usepackage{mlsys2025}
\hypersetup{pdfsubject={MLSys 2027}}
\fi

\newcommand{\ms}{\,ms}
\newcommand{\us}{\,\textmu s}

\mlsystitlerunning{Runway Characters: Real-Time Conversational Video at Batch Size One}

\begin{document}

\twocolumn[
\mlsystitle{Runway Characters: Serving a 57B Video Diffusion Transformer\texorpdfstring{\\}{ }for Real-Time Conversation}

\ifarxiv
\begin{mlsysauthorlist}
\mlsysauthor{Rohan Agarwal}{rw}
\mlsysauthor{Robin Andeer}{rw}
\mlsysauthor{Alex Armbruster}{rw}
\mlsysauthor{Bennett Arthur}{rw}
\mlsysauthor{Michail Christos Doukas}{rw}
\mlsysauthor{Piero Esposito}{rw}
\mlsysauthor{Qiming Fang}{rw}
\mlsysauthor{Aron Filkey}{rw}
\mlsysauthor{Bryan Fox}{rw}
\mlsysauthor{Anastasis Germanidis}{rw}
\mlsysauthor{Jonathan Granskog}{rw}
\mlsysauthor{Corina Gurau}{rw}
\mlsysauthor{Robin Kahlow}{rw}
\mlsysauthor{Taras Khakhulin}{rw}
\mlsysauthor{Nasir Mohammad Khalid}{rw}
\mlsysauthor{Tomasz \L{}akomy}{rw}
\mlsysauthor{Mike Leisz}{rw}
\mlsysauthor{Kathleen M. Lewis}{rw}
\mlsysauthor{Daniil Merkulov}{rw}
\mlsysauthor{Nicolas Neubert}{rw}
\mlsysauthor{Ryan Phillips}{rw}
\mlsysauthor{Yining Shi}{rw}
\mlsysauthor{Axel \"{O}rn Sigur\dh{}sson}{rw}
\mlsysauthor{Erica Simmons}{rw}
\mlsysauthor{Mark Sturley}{rw}
\mlsysauthor{Ziyi Tang}{rw}
\mlsysauthor{Michail Tarasiou}{rw}
\mlsysauthor{Jamie Umpherson}{rw}
\mlsysauthor{Cindy Wang}{rw}
\mlsysauthor{Jimei Yang}{rw}
\mlsysauthor{Hudson Yeo}{rw}
\mlsysauthor{Wei Zhang}{rw}
\mlsysauthor{Zejia Zheng}{rw}
\end{mlsysauthorlist}
\mlsysaffiliation{rw}{Runway, New York, NY, USA}
\mlsyscorrespondingauthor{Zejia Zheng}{zejia@runwayml.com}
\else
\begin{mlsysauthorlist}
\mlsysauthor{Anonymous Authors}{anon}
\end{mlsysauthorlist}
\mlsysaffiliation{anon}{Runway, New York, NY, USA}
\mlsyscorrespondingauthor{Anonymous Authors}{anon@example.com}
\fi

\mlsyskeywords{Video diffusion, real-time serving, KV cache, tensor parallelism, FP8}

\vskip 0.3in

\begin{abstract}
Conversational video characters must generate video as fast as it plays,
for one user at a time: at 24\,fps, a sustained 166.7\ms{} per four-frame
chunk at batch size one, with only a few chunks of buffering. We describe the
system behind Runway Characters, which serves a ${\approx}$57B-parameter
autoregressive video diffusion transformer (four denoising passes per chunk)
at 704$\times$1088 and 24\,fps on four H100 GPUs with a voice agent in the
loop. Each chunk takes 240 attention calls, each reading a 5,788-slot KV
cache in which only 757 tokens are new, and 29\% of the original step was
whole-cache work repeated in every call. We move that work to write time (a
ring cache, RoPE and FP8 quantization applied once per write, and a delta
rotation when positions reset), fuse the tensor-parallel all-reduce into the
FP8 GEMM with in-switch reduction, and give the decoder that turns latents
into video frames its own GPU. Without the
fused kernel, our optimizations cut the diffusion step from 240.9 to 159.3\ms{} (1.51$\times$,
24.5\,fps sustained) with no detected degradation in lip-sync scores;
the fused kernel, rolled back once in production and since re-enabled,
reaches 147.4\ms{}
(1.63$\times$, 43\% of FP8 peak). In seven weeks of production the worker
averaged above 23.5\,fps in 99.5\% of sessions that streamed for at
least a second, and our pipeline takes 0.49\,s of a 2.1\,s median
conversational turn. On the same GPUs, the two open-source real-time video
systems we benchmarked, at four denoising passes, miss real time or add over a
second of latency (one by estimate), at a third to half of our resolution.

\end{abstract}
]

\printAffiliationsAndNotice{}

\section{Introduction}
\label{sec:intro}

A conversational video character is a generated face that listens and talks
back in real time: the user speaks, a voice agent answers, and a video model
animates the character to that answer while it plays. Runway
Characters~\cite{runway2026characters} serves such characters with a
${\approx}$57B-parameter autoregressive (AR) video diffusion transformer.
Each user gets a private stream: every 4 video frames (166.7\ms{} at
24\,fps) the model must denoise one new latent frame (a compressed
frame that a separate decoder turns into the 4 video frames), conditioned on
recent audio and video, at that rate for the whole session. The player buffers only
a few chunks; falling behind by more freezes the picture, and since each frame
carries its slice of audio, the voice stalls with it.

We treat the model as a fixed workload. Runway Characters animates any single
image, from a photorealistic human to a cartoon mascot or fantasy creature, in
HD at 24\,fps~\cite{runway2026characters}, and we chose model capacity for
visual fidelity and temporal consistency across such characters in long,
high-resolution conversations. Our goal is to serve that model at
interactive rates in a deployed product without degrading its output.

Table~\ref{tab:requirements} lists what real time means here. It differs from
the serving problems the systems literature has studied most. LLM servers
amortize weights by batching requests; here the deadline is per user and per
chunk, and one session already saturates its GPUs. Video diffusion servers
parallelize long sequences offline~\cite{fang2024xdit}, or pipeline frames
and denoising steps across GPUs for
throughput~\cite{feng2026streamdiffusionv2,huang2025liveavatar}, which
multiplies per-chunk latency by the pipeline depth.

\begin{table}[t]
\caption{What real time means for a conversational video character. Lip sync:
detectability thresholds, $+$ = audio leads; turn gaps: means (agent: a commercial voice agent).}
\label{tab:requirements}
\centering
\footnotesize
\setlength{\tabcolsep}{4pt}
\begin{tabular}{@{}lll@{}}
\toprule
Requirement & Target / reference & Source \\
\midrule
Sustained rate & 166.7\ms{} per chunk & 24\,fps \\
Batch size & one session & compute-bound (\S\ref{sec:background}) \\
Lip-sync detection & +45 / $-$125\ms{} & \citealp{itu1998bt1359} \\
Human turn gap & 208\ms{} & \citealp{stivers2009turntaking} \\
Agent turn gap & 1.30\,s & \citealp{lin2025fullduplexbench} \\
\bottomrule
\end{tabular}
\end{table}

At batch size one, each AR step runs 4 denoising passes $\times$ 60 blocks
= 240 attention calls. Each call writes 757 new tokens into its own
5,788-slot KV cache and reads all of it, so work done per read touches
7.6$\times$ the tokens of work done per write, in every call. In the
original implementation, 29\% of the step was such work: re-applying rotary
position embeddings (RoPE) to the cached keys, re-quantizing the cache to FP8,
and shifting the cache window. The next cost after the GEMMs was the
tensor-parallel all-reduce, in series with every block. Attention itself was
6\% of the step.

\paragraph{Contributions.}
\begin{itemize}
\item A characterization of conversational video generation as a serving
workload: a sustained per-chunk rate at batch size one, 240 attention calls
per chunk that each re-read a mostly unchanged cache, and a turn budget shared
with a voice agent (\S\ref{sec:background}).
\item A write-time KV-cache design: a ring cache with a device-side write
position, RoPE and FP8 quantization applied once at write with per-(layer,
head) scales, and a delta rotation at position resets. It removes
91\% of cache upkeep (\S\ref{sec:writetime}).
\item A fused FP8 GEMM + all-reduce for odd tensor-parallel degrees on Hopper
and Blackwell that reduces output tiles in the NVLink switch while later
tiles are computed (\S\ref{sec:comm}).
\item The production system and its measurement: seven weeks of telemetry, a
per-stage turn breakdown, quality against a run-to-run floor, and comparisons
with open systems and Blackwell (\S\ref{sec:production}--\S\ref{sec:lessons}).
\end{itemize}

On four H100s our optimizations without the fused kernel run the diffusion
step in 159.3\ms{} (240.9\ms{} without them), which halved the GPUs
per session against the 8-GPU launch; the fused kernel, rolled back
once in production and since re-enabled, reaches 147.4\ms. On 53 clips and in 10-minute conversations
we detect no degradation in lip-sync scores. In production the worker averaged above
23.5\,fps in 99.5\% of sessions that streamed for at least a second, and of a 2.1\,s median turn our pipeline takes 0.49\,s.

\section{Background and Workload}
\label{sec:background}

\paragraph{Few-step AR video diffusion.}
The Runway Characters model generates one latent frame at a time, conditioned through a KV cache
on the frames before it. Sampling takes four denoising passes per latent
frame. A \emph{step} (one AR step) denoises one latent frame,
a four-frame chunk, in four forwards at batch size one. The model works on
compressed latent frames; a separate decoder network (the decoder of a
variational autoencoder, VAE) turns each denoised latent frame into 4 video
frames at 704$\times$1088, a step we call \emph{decoding}. The speech the character
says during those frames conditions the step.

\paragraph{The model.}
The model has 60 blocks. Under tensor parallelism (TP) each block's input
projections are column-parallel and its output projection row-parallel, so each
block needs one all-reduce. The
block linears, which run in every forward, hold ${\approx}$57B parameters and
run in FP8 with per-token activation and per-channel weight scales. The model keeps one KV cache per layer and denoising pass, so each of a
step's 240 attention calls writes and reads its own. Each has 1,246 global
slots, written once per session, and a window of 6 latent frames of 757 tokens:
5,788 slots, all read by attention. Temporal RoPE positions slide by one latent per step and reset every 42
steps, keeping them within a fixed range.

\paragraph{The step budget.}
One AR step is 240 block forwards over 757 tokens: ${\approx}$343\,TFLOP of
GEMMs and ${\approx}$30\,TFLOP of attention, or 93.3\,TFLOP per video frame of
useful work (attention over all 5,788 slots).
At 24\,fps a session sustains 2.24\,PFLOP/s, more than one H100's dense FP8
peak (1.98\,PFLOP/s). We estimate that three diffusion GPUs are the minimum
for our layout on H100: at TP-3 the FP8 GEMMs alone take 86\ms{}, and scaling
the whole TP-3 step by 3/2 gives ${\approx}$230\ms{} at TP-2, though
communication and fixed costs need not scale that way. We run batch size one
because the GEMMs already run near two-thirds of FP8 peak at 757 tokens, so a
second session in the batch would nearly double the step.

\begin{table}[t]
\caption{GPU time per AR step on one diffusion GPU (4$\times$H100, TP-3; 10
profiled steps of a live-length session; ms). Cache upkeep: whole-cache RoPE,
padding, window shifts and FP8 re-quantization. Fused: the output projection's
GEMM and all-reduce run as one kernel; FP8 GEMMs is the remaining GEMMs.}
\label{tab:profile}
\centering
\footnotesize
\setlength{\tabcolsep}{4pt}
\begin{tabular}{@{}lrrr@{}}
\toprule
Component & Baseline & Prod. & Fused \\
\midrule
FP8 GEMMs & 84.4 & 85.9 & 54.1 \\
Fused GEMM + all-reduce & -- & -- & 40.9 \\
KV-cache upkeep & 69.7 & 6.2 & 6.4 \\
TP all-reduce (NCCL) & 28.6 & 21.2 & -- \\
Activation quantization & 19.7 & 16.9 & 18.5 \\
Attention kernel & 13.7 & 14.1 & 15.0 \\
AdaLN modulation & 11.8 & 0.8 & 0.7 \\
Other & 11.4 & 8.8 & 3.4 \\
\midrule
GPU busy & 239.2 & 153.9 & 138.9 \\
\bottomrule
\end{tabular}
\end{table}

\paragraph{Where the original step went.}
In the baseline (Table~\ref{tab:profile}) the GEMMs already ran at
${\approx}$67\% of FP8 peak. Cache upkeep took 69.7 of 239.2\ms{} (29\%):
keys were cached before RoPE, so every block of every denoising pass padded
the queries to the cache length and re-applied RoPE to all cached keys; FP8
attention re-quantized the whole cache in each of the 240 calls; and the
window advanced by copying it. The all-reduce (28.6\ms) followed each
down-projection GEMM. The attention kernel took 13.7\ms.

\section{System Overview}
\label{sec:system}

\begin{figure}[t]
\centering
\includegraphics[width=\columnwidth]{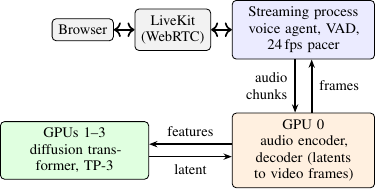}

\vspace{6pt}
\includegraphics[width=0.95\columnwidth]{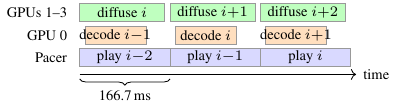}
\caption{Top: one session on four H100 GPUs. Bottom: in steady state three
steps are in flight; decoding of step $i{-}1$ into video frames and playback of step $i{-}2$
overlap the denoising of step $i$.}
\label{fig:arch}
\end{figure}

A session runs on four H100 GPUs (Figure~\ref{fig:arch}). The browser
connects over WebRTC to a LiveKit~\cite{livekit2026livekit} room. A streaming process
hosts the connection to a third-party voice agent, a local
voice-activity detector for idle handling, and the pacer that publishes
frames and audio at 24\,fps. GPU 0 runs the audio encoder and the decoder
(latents to video frames) and broadcasts each step's audio features; GPUs 1--3 run the
transformer with TP-3 and send the denoised latent back.

\paragraph{Per-step schedule.}
Three steps are in flight: while GPUs 1--3 denoise latent $i$, GPU 0 decodes
latent $i{-}1$ (114\ms) and the pacer plays latent $i{-}2$. GPU 0 encodes the
audio for step $i$ (9\ms) and broadcasts it before decoding latent $i{-}1$, so
in a live session latent $i{-}1$ is decoded only once audio $i$ has arrived.
Decode overlaps diffusion fully, leaving a serial remainder of 1.6\ms{} per
step. Every queue has depth one with backpressure. With the queued audio, denoising,
decoding and the playout buffer, 3--5 chunks of speech are in flight, which
sets the latency of a turn and of an interruption (\S\ref{sec:eval-latency}). On an underrun the pacer holds the last frame and
shifts its schedule by the delay rather than dropping frames, and the audio
inside the frames waits with it.

\paragraph{Why tensor parallelism and a separate decoder GPU.}
Pipelining layers or denoising passes across $G$ GPUs, as
StreamDiffusionV2~\cite{feng2026streamdiffusionv2} and Live
Avatar~\cite{huang2025liveavatar} do, puts about $G\times$ steps pipeline
ticks between an audio chunk and its frames: an estimated 3.2\,s for
StreamDiffusionV2 at four steps on four H100s (\S\ref{sec:eval-oss}). Sequence
parallelism~\cite{fang2024usp} would split only 757 query tokens. Tensor
parallelism shortens each forward, at one all-reduce per block. Decoding on a
diffusion GPU would add its 114\ms{} to every step, since the TP group waits
for its slowest member; on B200 we measured that it misses real time
(\S\ref{sec:eval-scaling}).

\section{Moving Work from Read Time to Write Time}
\label{sec:writetime}

\begin{figure}[t]
\centering
\includegraphics[width=\columnwidth]{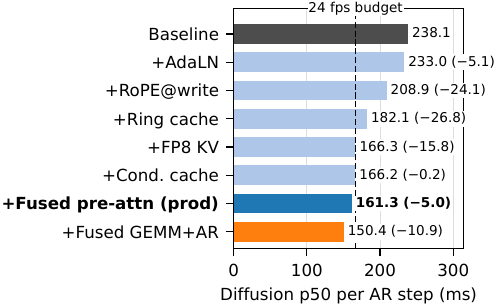}
\caption{Cumulative waterfall on one 4$\times$H100 box (TP-3): diffusion time
per AR step and the change from the previous configuration: medians over three
runs per arm in random order, each a fresh worker and one 180\,s session; every
arm's three values lie within 1.0\ms. The fused GEMM +
all-reduce is \S\ref{sec:comm}.}
\label{fig:waterfall}
\end{figure}

The common thread of this section is to do once per write what was done once
per read. Figure~\ref{fig:waterfall} gives the effect of each change, applied
cumulatively in the order the code's dependencies allow.

\subsection{A ring cache with a device-side write position}
In the original code (the baseline of Table~\ref{tab:profile}), the cache was
one fixed-size allocation per layer and denoising pass, and once full it shifted the whole window left with two overlapping copies per
step. We stack all layers into one allocation, keep per-layer views, and write
each new frame into a circular slot with \texttt{index\_copy\_}. The write
position is a device tensor, so the compiled step and its CUDA graph capture a
pointer rather than a Python integer and replay for every step.

A ring is correct only if each cached frame keeps the position it was written
with. The original code re-applied RoPE in physical slot order, which stops
being chronological once the ring wraps: combined with the ring it silently
permutes the temporal positions of cached frames. A test now pins that
combination as invalid.

\subsection{RoPE at write, and a delta rotation at reset}
We apply RoPE to each new token's queries and keys once, before the cache
write, and attention consumes cached keys as they are. This removes the
query padding, the full-cache RoPE and the full-length concatenations, about
0.5\,GB of memory traffic per block forward.

Temporal positions cycle within a fixed range, and every 42 steps
every cached frame's temporal index moves by the same $\Delta = -42$. The
temporal angle of a window frame is affine in its index $t$, $\theta_j(t) =
\omega_j (a t + b)$, so the reset is one rotation of each cached window key by
$\omega_j a \Delta$ per frequency $j$; spatial angles and global tokens do not
change. In an FP8 cache the key is dequantized, rotated in fp32 and
requantized with its unchanged per-head scale: the rotation identity is
exact, and the stored key changes by one FP8 rounding, at most once per key
(Appendix~\ref{app:reset}). A reset costs one pass
over the window every 42 steps instead of a full-cache RoPE in each of 240
calls per step. It shows up as one slower step in 42 (263\ms{} in the
production configuration, \S\ref{sec:eval-delivery}), which the playout
buffer absorbs.

\subsection{FP8 at write}
FP8 attention originally recomputed a per-tensor scale and re-quantized the
whole cached K and V in every call. We store K and V in \texttt{float8\_e4m3}
and quantize only the new frame, with per-(layer, head) scales: the descale
granularity that FlashAttention-3's FP8 kernels take as
arguments~\cite{shah2024flashattention3,dao2026fa3code}, so they need no
separate rescaling kernel. Only the queries are quantized
per call.

Scales are calibrated once per session, from the amax of the bf16 cache
filled by the first six steps, times a margin of 1.5; values beyond it
saturate. Because K and V are projections of RMS-normalized tokens, input
gain cannot move their amax, but content can. Rare clipping alone does not
bound attention error, so \S\ref{sec:eval-quality} compares with a bf16 cache;
a rolling amax cost a step (\S\ref{sec:lessons}). Scales are per head because
value heads differ widely in range: one per-tensor scale would leave the
quietest heads few significant bits.

\subsection{Fused pre-attention}
The remaining work between the up-projection and attention is a chain of small
memory-bound operations: split, reshape, RMS-normalize Q and K, RoPE, quantize
K and V per head, write them into the ring. Three CuTe DSL kernels (Q, K, V)
fuse it; the K and V kernels read the write position from device memory. The
fused RMS norm squares in fp32 where the reference squares in bf16, which
flips the last bf16 bit of about 2\% of elements. At RoPE cancellation points
($x_0\cos\theta \approx x_1\sin\theta$) that bit dominates, so 0.3\% of FP8
codes differ by a few codes while the absolute error stays within two bf16
ulps of the head scale; tests therefore bound absolute, not relative, error.

\subsection{Constant folding over a fixed schedule}
The four-pass schedule is fixed, so all 60 blocks' AdaLN
modulations are computed in one batched GEMM outside the compiled region and
cached across steps (11.8 $\rightarrow$ 0.8\ms). The four forwards of a step
share their conditioning, so RoPE tables and conditioning embeddings are
built once per step.

\paragraph{Effect.}
The three cache changes save 66.7 of the 76.7\ms{} the production
configuration saves, and cut cache upkeep from 68 to 6\ms. The conditioning
cache does not move the step: the host work it removes was already hidden
behind GPU work. RoPE at write is also what introduces the reset step; the
baseline re-applies RoPE every step and has none.

\section{Hiding Tensor-Parallel Communication}
\label{sec:comm}

With the cache work gone, the all-reduce after each down-projection was the
largest cost after the GEMMs: at TP-3 the FP8 down-projection followed by an
NCCL all-reduce of its output takes 208\us, 240 times per step.

\paragraph{Swapping the collective did not help.}
Replacing NCCL with an NVLink SHARP (NVLS) multimem all-reduce, which offloads
reduction and multicast to the NVSwitch, saved 0.4\ms{} per step. The FP8 GEMM could not write
into the symmetric buffer, forcing a copy, and the reduction still started
only after the whole GEMM.

\paragraph{Fusing the reduction into the GEMM.}
We extend the fused GEMM + all-reduce of
ParallelKittens~\cite{sul2026parallelkittens}, built on
ThunderKittens~\cite{spector2025thunderkittens}. A persistent kernel fills
all 132 SMs. Compute SMs write each output tile into their symmetric-memory
buffer and bump a counter at the tile's owner rank; once all $N$ partials
have landed, the owner reduces the tile in the switch with
\texttt{multimem.ld\_reduce} and writes the sum to every rank with
\texttt{multimem.st} (Figure~\ref{fig:kernelproto}). Tiles are reduced while
later tiles are still being computed, so most communication leaves the critical path. Four changes to the
upstream kernel were needed:
\begin{enumerate}
\item \textbf{FP8 GEMM.} Upstream's Hopper kernel is bf16, and a bf16
down-projection costs more than the whole all-reduce (+8.3\ms{} per step).
Ours takes FP8 inputs with per-token and per-channel scales,
accumulates in fp32 and reduces in bf16, since \texttt{multimem.ld\_reduce}
has no FP8 form on Hopper.
\item \textbf{Vectorized reduction.} Upstream moved 4 bytes per thread per
round trip and needed 32 of 132 SMs to keep pace; we use the 16-byte
\texttt{ld\_reduce.v4.bf16x2} form with two accesses in flight.
\item \textbf{Shared tail queue.} With a static compute/communication split
the tiles of the last GEMM wave, finishing together, waited on a few
communication SMs (34 of 176\us). Compute SMs now drain into a shared
reduction queue, which made the split nearly irrelevant: 2 communication SMs
give 153.0\us, 8 give 155.4\us; 16 or more starve the GEMM (184--187\us).
\item \textbf{Odd world sizes.} TP-3 and TP-7 groups live on devices
1..$N$, so the world size is a template parameter and symmetric memory comes
from PyTorch rather than upstream's handshake, which assumes rank equals
device index.
\end{enumerate}

\begin{figure}[t]
\centering
\includegraphics[width=\columnwidth]{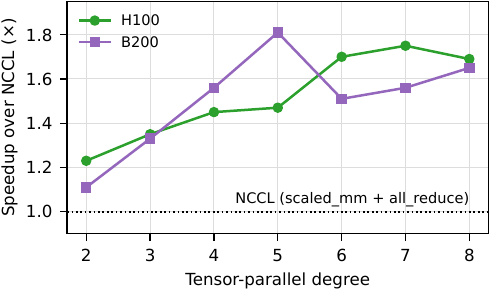}
\caption{Fused GEMM + all-reduce vs FP8 GEMM then NCCL all-reduce at the
production shape ($M{=}757$, each rank's slice of the down-projection;
Appendix~\ref{app:shapes}). H100 TP-2
to TP-4 were measured on a 4-GPU box, TP-5 to TP-8 on an 8-GPU node.}
\label{fig:kernel}
\end{figure}

\paragraph{Microbenchmarks.}
At the production shape ($M{=}757$ tokens, TP-3) the fused
kernel takes 153.8\us{} against 208.2\us{} for \texttt{\_scaled\_mm} + NCCL.
Its own GEMM is 5--22\% slower than \texttt{\_scaled\_mm}; in five runs timing
both alone, the fused call hides 54--68\% of the all-reduce's time. Fusion needs a GEMM large
enough to hide the reduction under and a reduction large enough to matter:
fused vs NCCL is 0.84$\times$ at $M{=}64$, 1.35$\times$ at 757 and
1.01$\times$ at 11,136. The benefit grows with the TP degree
(Figure~\ref{fig:kernel}), as the ring's traffic grows and the GEMM shrinks.

\paragraph{Blackwell.}
The Blackwell kernel, which we run on B200, rebuilds the GEMM on tcgen05 MMAs
with 2-CTA clusters and 256$\times$256 tiles in tensor memory and keeps the
Hopper reduction, at a finer grain: reducing whole tiles left most compute
clusters idle while the last wave was reduced, so it reduces 128$\times$64
sub-tiles. B200's FP8 GEMM is 1.9$\times$ faster than H100's but its GEMM + NCCL
call only 1.76$\times$, so at TP-2 there is less GEMM to hide the reduction under
(1.11$\times$); from TP-4 up the gain is 1.5--1.8$\times$
(Blackwell pitfalls: Appendix~\ref{app:kernel}).

\paragraph{Correctness under compilation.}
Two bugs passed every eager test and failed only when compiled. The operator
wrote the symmetric buffer without declaring it (declaring it lets
functionalization substitute a clone in ordinary memory, where
\texttt{multimem} silently does not reduce), and all calls at one shape share
the buffer, so Inductor legally moved one call's read past the next call's
write: 96\% of elements wrong with 60 calls in a graph. The operator now
returns a copy of the result, at no measurable cost. And CUDA graph trees copy
unmarked symmetric buffers into their own pool, so the captured kernel wrote
partials its peers never read; the buffers must be marked as static.

\paragraph{Status.}
The kernel served production in August 2026 until, after a PyTorch and NCCL upgrade, workers ran
out of memory binding NVLS multicast buffers; reverting to NCCL was a
configuration change (\S\ref{sec:lessons}). The kernel kept buffers for every
token count, and a session's first forward has a prompt-dependent length, so
workers bound new buffers per prompt length, which fixed-prompt benchmarks
never do. Now only the steady-state shape is fused and first forwards use
NCCL. After a 15-hour canary in which every session with a measured worker fps
averaged 24.0\,fps or more, the kernel was re-enabled on 8 October 2026. In
the 15 days since, 0.7\% of sessions of at least a second averaged 24\,fps or
less, against 7--23\% in each of the five weeks before, and 0.08\% averaged
23.5\,fps or less.

\section{Operating It in Production}
\label{sec:production}

\paragraph{Deployment.}
Runway Characters launched on eight H100s per session and then served on
four B200s, both without the optimizations of \S\ref{sec:writetime}; with
them it serves on four H100s, half the launch configuration.
Workers pull sessions from a queue and serve one at a time, so capacity is
the number of warm workers.

\paragraph{Worker start-up.}
A worker takes a median 6.5\,min from pod creation to ready, 4.2\,min of it
compiling and warming up, much longer than a typical session, so capacity
cannot follow demand per session and must be held warm.

\paragraph{Session start.}
A new session resets diffusion state, not the worker: it encodes the new
reference image and prompt and clears the dynamic cache (0.4--0.6\,s), while
compiled graphs and static buffers persist. The first six steps run eagerly
while the cache fills (4.0--4.5\,s). In live sessions frames were released
5.4--9.9\,s after the worker began the task (queue time excluded).

\paragraph{Health.}
The worker measures its own fps and logs every session that streams below
target. A policy that replaces a worker after two consecutive slow sessions is
implemented but switched off until its decisions have been checked against
production logs.

\section{Evaluation}
\label{sec:eval}

\paragraph{Setup.}
Unless noted, we use one 4$\times$H100 SXM box with NVSwitch, PyTorch 2.13
(CUDA 13.0) and NCCL 2.29.7, in the production layout. A benchmark harness
drives the production worker's components in production order (setup and
warm-up, then per session the task reset and an AR step loop fed from recorded
speech) without the queue, LiveKit or voice agent. It runs unpaced unless
stated, so its step period is capacity. Each arm is a fresh worker; sessions
are 300\,s (${\approx}$1,800 steps) with the window full and resets included,
and steady state excludes the first 60 and the profiled steps.
\emph{Baseline} turns every optimization off; it is the launch software,
which served on eight GPUs per session, replayed here on four. \emph{Production} is the set deployed while we
measured, and \emph{fused} adds the fused GEMM + all-reduce. We also use
live sessions on the staging app and seven weeks of production telemetry.
Production results use performance telemetry only: aggregate frame-rate and
timing counters and per-session frame-rate records; no user audio, video,
images or text were accessed. Live sessions and
latency runs were the authors' own, driven by synthesized speech.

\subsection{Real-time delivery}
\label{sec:eval-delivery}

\begin{table}[t]
\caption{Live-length benchmark sessions (2 $\times$ 300\,s per arm, mean of
the two). Capacity is $4000/\text{period p50}$; sustained is frames generated
over elapsed time, resets included.}
\label{tab:e1}
\centering
\footnotesize
\setlength{\tabcolsep}{4pt}
\begin{tabular}{@{}lrrr@{}}
\toprule
Per AR step & Baseline & Prod. & Fused \\
\midrule
Diffusion p50 (ms) & 240.9 & 159.3 & 147.4 \\
Period p50 (ms) & 242.8 & 161.0 & 149.1 \\
Period p90 (ms) & 245.2 & 163.4 & 151.7 \\
Capacity (fps) & 16.5 & 24.8 & 26.8 \\
Sustained (fps) & 16.4 & 24.5 & 26.4 \\
Reset step, every 42 (ms) & -- & 263 & 245 \\
TFLOP/s per diffusion GPU & 517 & 781 & 844 \\
\% of dense FP8 peak & 26.1 & 39.5 & 42.7 \\
\bottomrule
\end{tabular}
\end{table}

\begin{figure*}[t]
\centering
\includegraphics[width=\textwidth]{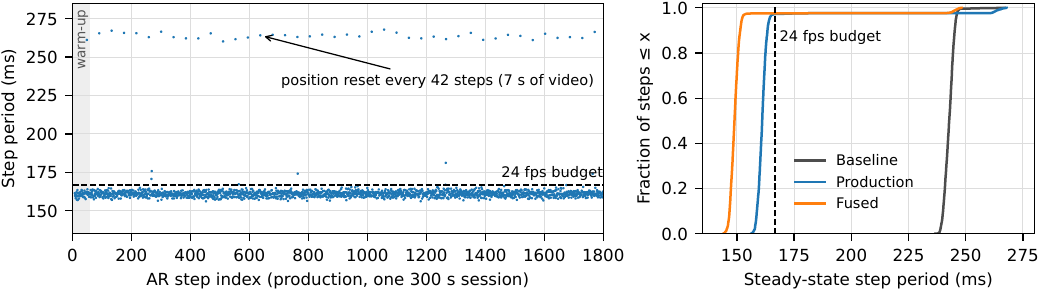}
\caption{Left: step period over one 300\,s production-configuration session;
the spikes are position resets every 42 steps. Right: distribution of the
steady-state period; the dashed line is the 24\,fps budget.}
\label{fig:steptrace}
\end{figure*}

\paragraph{Benchmark.}
The production configuration runs the diffusion step in 159.3\ms, 1.51$\times$
faster than the baseline, which cannot run in real time on this layout; the
fused kernel reaches 1.63$\times$ (Table~\ref{tab:e1},
Figure~\ref{fig:steptrace}). The period exceeds diffusion by only 1.6\ms,
because decoding and audio encoding overlap it. The steps over budget are the
resets (2.4\%) and a few of 167--189\ms{} (0.4\%); with them the
configuration sustains 24.5\,fps, 2\% above real time. The contract is
therefore a sustained rate with bounded excursions: a reset runs 96\ms{} over
budget, and a simulated 24\,fps player sees no stalls when generation may lead
playback by two steps, one finished step queued (with one, each reset stalls
it for ${\approx}$100\ms; Appendix~\ref{app:slack}). Live sessions on stage matched the benchmark within 2\ms{} of
diffusion time, and the browser showed 24\,fps with no dropped frames.

\paragraph{Production.}
We report 4$\times$H100 workers from 16 August to 4 October 2026 (the fused
kernel on until 31 August; Table~\ref{tab:prod}). While streaming, the worker reports its fps per 10\,s
interval and every fifth frame's drift from its slot on the pacer's clock;
after a late frame the pacer shifts its clock, so a stall counts once, at its length. The interval in which a stream
ends also counts its idle tail, so intervals overstate slow generation (in
three hours of logs: 2.2\% at 23.5\,fps or less, against 0.3\% of logged windows cut
at the last frame).
Of the sessions
that streamed at least a second, 0.55\% averaged 23.5\,fps or less on the
worker. 0.16\% of sampled frames
(sampling meets resets at every phase) were more than 10\ms{} off their slot,
a quarter of the 0.6\% that a stall at every reset would cause, so the
playout buffer absorbs most resets in production too.

\begin{table}[t]
\caption{Production, 4$\times$H100 workers, 16 Aug -- 4 Oct 2026: per-series
sums of observed Prometheus counters (Appendix~\ref{app:repro}); fractions at
bucket edges.}
\label{tab:prod}
\centering
\footnotesize
\setlength{\tabcolsep}{4pt}
\begin{tabular}{@{}lr@{}}
\toprule
Sessions ${\ge}$1\,s, worker avg.\ $\le$23.5\,fps & 0.55\% \\
Report intervals $\le$23.5 / $\le$20\,fps & 2.3\% / 0.8\% \\
Frames (1 in 5) within 1/10/100\ms{} of slot & 99.52/99.84/99.98\% \\
\bottomrule
\end{tabular}
\end{table}

\subsection{Latency}
\label{sec:eval-latency}
\begin{figure}[t]
\centering
\includegraphics[width=\columnwidth]{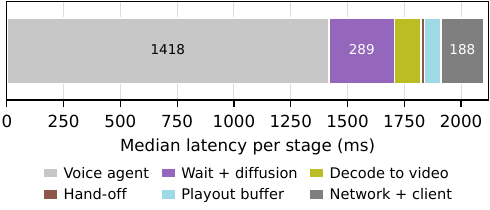}
\caption{Per-stage median turn latency, production configuration (46 turns,
11 live sessions; 43 for stages from the user's end of speech), from the user's end of speech to the avatar's audio at the
client. Stage medians need not sum to the median total.}
\label{fig:turn}
\end{figure}

We joined server traces with the browser's trace for 56 turns in 12 live
sessions and excluded five whose first speaking frame fell in the profiler's
window; of the other 51, 46 (11 sessions) ran the production configuration
and five (one session) the fused kernel (Figure~\ref{fig:turn}). Server
stages and client totals are each on one clock; only the split between the
first and last stage depends on the clock-sync offset (${\pm}$100\ms). The
median turn, from the user's end of speech to the avatar's audio at the client,
is 2.13\,s (p10--p90 1.45--3.00\,s) and the voice agent takes 1.42\,s, over
the 43 turns with a user end-of-speech mark. Our
pipeline takes 493\ms{} from the agent's first audio to the first speaking
frame captured by LiveKit (p10--p90 421--562\ms); the stage medians are
289\ms{} to the denoised latent, waiting for the in-flight step included, 117\ms{} decoding,
14\ms{} hand-off and 74\ms{} in the playout buffer. Delivery takes another ${\approx}$190\ms. The fused
kernel does not shorten the turn: under pacing the leader decodes latent $i$
only after audio $i{+}1$ arrives, so saved diffusion time becomes waiting. It
buys capacity, not latency. When the user interrupts, the 3--5 voiced latents
already in flight still play, and the last voiced frame reaches LiveKit on
the server a median 0.78\,s after the agent's interruption signal
(0.70--1.01\,s, 11 cases), not timed at the client.

\subsection{Quality}
\label{sec:eval-quality}
Three optimizations change numerics: the FP8 cache, the fused pre-attention
and the fused all-reduce. We generated all 53 clips of our evaluation set (50
reference images in photoreal, cartoon, claymation, fantasy and non-human
styles; 954\,s of speech) with baseline, production and fused, with production
four times to measure run-to-run variation, and with a bf16 cache to isolate
the FP8 cache. Lip sync is SyncNet~\cite{chung2016syncnet} distance and confidence
(LSE-D, LSE-C) with Euclidean distance on video resampled to 25\,fps, as in
the standard protocol~\cite{prajwal2020wav2lip} but with our own crop.
SyncNet does not track every style, so arms are compared on the 51 clips it
scores in every run (Appendix~\ref{app:quality}). We detected no degradation
in lip sync (Table~\ref{tab:quality}). Mean differences are at most 0.03 and
every 95\% interval lies within $\pm$0.08; the four production runs span
0.024 LSE-D and 0.018 LSE-C, so differences of this size also occur between
repeated runs. SyncNet is weaker on the 15 stylized clips it tracks (LSE-C
5.5, photoreal 8.1), with intervals 2--5$\times$ as wide. The
intervals bound the changes but do not prove equivalence.

\begin{table}[t]
\caption{Lip sync: paired per-clip differences (mean, 95\% bootstrap
interval) over the 51 clips SyncNet scores in every run. Absolute values:
LSE-D 8.26--8.29 (lower is better), LSE-C 7.32--7.36. Last row: each of three
reruns.}
\label{tab:quality}
\centering
\footnotesize
\setlength{\tabcolsep}{4pt}
\begin{tabular}{@{}lrr@{}}
\toprule
& LSE-D & LSE-C \\
\midrule
Prod.$-$base & $-$.02 {\scriptsize[$-$.07,.02]} & +.03 {\scriptsize[$-$.01,.07]} \\
Fused$-$base & .00 {\scriptsize[$-$.04,.04]} & +.01 {\scriptsize[$-$.03,.06]} \\
bf16 KV$-$prod. & +.02 {\scriptsize[$-$.02,.06]} & $-$.01 {\scriptsize[$-$.05,.03]} \\
Reruns$-$prod. & $-$.01/+.01/+.02 & +.01/$-$.01/$-$.01 \\
\bottomrule
\end{tabular}
\end{table}

\paragraph{Long sessions.}
The clips are short (median 18.5\,s, at most 40\,s), while production sessions
can last much longer. As a targeted stress test we ran four 10-minute
sessions per arm that alternate speaking and listening as the worker does
(Appendix~\ref{app:quality}): a photoreal and a non-human character, each calibrated once while
listening and once while speaking, for production, a bf16 cache and a
production rerun on identical inputs. Over 130 speaking periods the bf16
cache changes LSE-D by +0.02 [$-$0.10, +0.14] and the rerun by +0.05 [$-$0.05,
+0.15]. These intervals resample segments, not sessions, so they understate
the uncertainty. Per session (Appendix~\ref{app:quality}) LSE-D moves as much
with the rerun as with the bf16 cache, and neither difference grows from the
first two minutes to the last beyond the rerun's own variation; four sessions
cannot exclude a trend.

\paragraph{Pixels stop separating runs after seconds.}
An AR model turns any numeric difference, even nondeterminism, into a
different trajectory: after 10\,s two production runs differ in PSNR as much
as any two configurations (Appendix~\ref{app:pixels}), so an optimization must
be judged on task metrics against a measured run-to-run floor.

\subsection{Scaling and layout}
\label{sec:eval-scaling}
Production packs two 4-GPU workers per 8-GPU node; two fused workers started
together on one node ran diffusion at 148.6 and 148.1\ms{} (147.4\ms{} alone).
On eight H100s (decoder + TP-7, heads split unevenly;
Appendix~\ref{app:shapes}) diffusion falls to
118.2\ms{} (98.9\ms{} fused), but the period stops at 128.7\ms{} (31\,fps): the single decoder GPU
becomes the bottleneck, and the fused kernel's 19\ms{} does not raise
throughput. A channels-last decoder (cuDNN's native NDHWC kernels) cuts
decoding from 117 to 96\ms, lifting the 8-GPU period to 119.4 and 109.0\ms{} fused. Sharing a GPU between decoding and
diffusion misses real time on B200: with the fused kernel, two B200s take 195.0\ms{}
even with an FP8 decoder, against 148.8\ms{} on three with a dedicated bf16
decoder. The shared GPU
stays at its 1000\,W power cap, slowing its diffusion kernels by
18--26\% and stalling the other rank in every all-reduce.

\subsection{Open-source systems on the same hardware}
\label{sec:eval-oss}
We ran StreamDiffusionV2~\cite{feng2026streamdiffusionv2} and Live
Avatar~\cite{huang2025liveavatar}, 14B models built on Wan~2.1 and Wan~2.2-S2V, from their public
code on the same box (Table~\ref{tab:oss}; recipes in Appendix~\ref{app:recipes}).
At four steps StreamDiffusionV2's
four-stage pipeline either misses real time (480$\times$832) or adds an
estimated 2.2\,s of latency (512$^2$), since a chunk crosses 16.5 pipeline ticks. Live Avatar's
released recipe needs five GPUs; on four it reaches 21.3\,fps, and a 3+1 layout
reaches 42.3\,fps only with a three-step schedule it was not distilled for, at
1.1\,s latency. At
704$\times$1088 the 14B model needs 166\,TFLOP per frame against our 93 and
runs at 6.3\,fps. These are feasibility comparisons of complete systems on
four GPUs, our production unit: model, tokenizer, resolution, schedule and
layout all differ, and we did not compare output quality, so the case for
our techniques rests on the ablations above.

\begin{table*}[t]
\caption{Real-time video systems on one 4$\times$H100 box. fps is capacity
(unpaced). Latency (p50): a chunk's arrival to its frames on the host, with
arrivals paced at the fps in parentheses (u: unpaced, from rank 0's pickup). TFLOP/s per GPU: model TFLOP per frame
$\times$ capacity fps $\div$ 4. PP: pipeline over layers; TPP: pipeline over denoising passes; + dec.: a GPU of its own decodes latents to video frames; shared dec.: decoding shares a diffusion GPU. $^\dagger$Re-enabled in production on 8 October 2026. $^\ddagger$Index-matched; alignment of the decoded output unvalidated.}
\label{tab:oss}
\centering
\footnotesize
\setlength{\tabcolsep}{3pt}
\begin{tabular}{@{}lrrrlrrrr@{}}
\toprule
System & Params & Resolution & Steps & Layout & fps & Real time & Latency & TFLOP/s/GPU \\
\midrule
\textbf{Runway Characters, prod.} & 57B & 704$\times$1088 & 4 & TP-3 + dec. & 24.8 & yes & 293\ms{} (24) & 579 \\
\textbf{Runway Characters, fused$^\dagger$} & 57B & 704$\times$1088 & 4 & TP-3 + dec. & 26.8 & yes & 292\ms{} (24) & 625 \\
StreamDiffusionV2 & 14B & 480$\times$832 & 4 & PP-4 & 20.6 & no & 3.20\,s (u)$^\ddagger$ & 257 \\
StreamDiffusionV2 & 14B & 512$\times$512 & 4 & PP-4 & 30.2 & yes & 2.18\,s (u)$^\ddagger$ & 227 \\
StreamDiffusionV2 & 14B & 512$\times$512 & 1 & PP-4 & 57.2 & yes & 0.34\,s (u)$^\ddagger$ & 108 \\
Live Avatar (released) & 14B & 384$\times$704 & 4 & TPP-4, shared dec. & 21.3 & no & 1.43\,s (20) & 205 \\
Live Avatar (3-step) & 14B & 384$\times$704 & 3 & TPP-3 + dec. & 42.3 & yes & 1.13\,s (25) & 305 \\
Live Avatar & 14B & 704$\times$1088 & 4 & TPP-4, shared dec. & 6.3 & no & -- & 263 \\
\bottomrule
\end{tabular}
\end{table*}

\subsection{End-to-end latency}
\label{sec:eval-e2e}
Published figures cover different intervals (Appendix~\ref{app:e2e}). To
measure Runway Characters through its public API we built a harness around a LiveKit Agents~\cite{livekit2026agents} worker that
sends 30 recorded agent responses (about 3, 8 and 15\,s) to an avatar through
its LiveKit plugin, while a viewer on the same host and clock times the
avatar's first voiced audio (Appendix~\ref{app:harness}); a mock avatar with a
known delay shows the harness adds 112\ms. Runway Characters takes 0.56\,s at
the median (p10--p90 0.46--0.78\,s, including the 112\ms), close to the
0.68\,s sum of the live sessions' stage medians from agent audio to client
audio (493 + ${\approx}$190\ms, \S\ref{sec:eval-latency}). The phase at which
agent audio meets the 166.7\ms{} step contributes to the spread.
We report our public-API measurement and local open-system benchmarks; other
commercial avatar APIs are not included. The open Live Avatar
encodes its whole audio file up front, so as released it cannot be driven
live. Adding the 112\ms{} to its paced latency on five H100s (1.41\,s from a
block's scheduled audio arrival to its frames on the host) gives an estimate
of 1.52\,s if transport simply adds; for our system the same sum falls
150\ms{} short of the measurement.

\subsection{Blackwell}
\label{sec:eval-b200}
On an 8$\times$B200 node with the same stack, three B200s run production
precision in real time: 154.3\ms{} per step (decoder + TP-2), 142.9\ms{} with
the fused kernel and a bf16 cache; with those, four take 107.4\ms{} and eight,
with an FP8 decoder too, 65.5\ms. Two
H100 optimizations behave differently. The FP8 cache saves 17.0\ms{} on H100
but costs 2.4 to 6.7\ms{} on B200, where our build has no FP8 attention that
takes its per-head scales, so the cache is dequantized whole before every call.
cuDNN's MXFP8 attention does not fit per-write quantization
(Appendix~\ref{app:reset}). The fused kernel matters most
where decoding is off the critical path: 1.20$\times$ end to end at TP-7. With
NVFP4 matmuls~\cite{alvarez2025nvfp4} and FP8 decoding two B200s reach real time (141.4\ms), but NVFP4
output sits ${\approx}$3\,dB below the run-to-run PSNR floor, so we treat it as
a speed bound.

\section{Lessons}
\label{sec:lessons}

\paragraph{Profile the current build.}
The multimem all-reduce was sized from an old trace under a heavy
profiler (${\approx}$150\us{} per all-reduce); on the current build it was
76\us, partly waiting on the slowest rank; only fusion recovered it.

\paragraph{Test custom collectives under the compiler.}
Both serious fused-kernel bugs (\S\ref{sec:comm}) were invisible in eager
mode, where issue order is execution order. Every kernel test now chains
calls under \texttt{torch.compile} with CUDA graphs.

\paragraph{One synchronization per step costs a step.}
One \texttt{.item()} per step, to check a rolling FP8 amax, took the step from
168.5 to 278\ms{} by stopping the CPU from running ahead of the GPU.

\paragraph{Keep the plain path behind every switch.}
The fused kernel worked for weeks, then failed in production on input variety
our benchmarks lacked (\S\ref{sec:comm}). Every optimization has a switch in one
configuration object that keeps the plain path, so turning it off was a
configuration change.

\paragraph{Measure the measurement.}
Two instruments misled us until we cross-checked them. GPU spans timed from
a CUDA event recorded at worker start had drifted 58--65\ms{} from host time
hours later, so we re-anchor them to host events on the same rank. And the
browser's ``avatar audio ended'' marker follows turn messages, not
sound, so barge-in latency starts from the agent's interruption signal.

\section{Related Work}
\label{sec:related}

\paragraph{Real-time autoregressive video diffusion.}
CausVid~\cite{yin2025causvid} and Self Forcing~\cite{huang2025selfforcing}
distill bidirectional video diffusion into few-step causal generators that
extend video through a KV cache. LongLive~\cite{yang2026longlive}, Rolling
Forcing~\cite{liu2026rollingforcing}, Deep Forcing~\cite{yi2026deepforcing} and
Infinity-RoPE~\cite{yesiltepe2026infinityrope} reach long horizons with
attention sinks, cache recaching and positional schemes; Deep Forcing and
Infinity-RoPE re-position cached keys as the window moves, and LongLive's
later relative-RoPE code~\cite{nvlabs2026longlivecode} stores unrotated keys
and applies RoPE at every read, work we move to write time. LongLive-2.0~\cite{chen2026longlive2} and Matrix-Game
3.0~\cite{wang2026matrixgame3} also give decoding its own GPU, and Quant
VideoGen~\cite{xi2026quantvideogen} and Forcing-KV~\cite{ji2026forcingkv}
quantize AR video caches. Seaweed APT2~\cite{lin2025apt2} (8B, eight H100s, 720p,
24\,fps), MAGI-1~\cite{sandai2025magi1} (24B, 24 H100s) and Krea
Realtime~\cite{millon2025krearealtime} (14B, one B200, 11\,fps) stream video.
These works improve models and schedules; we serve a larger model under a
conversational deadline.

\paragraph{Real-time talking avatars.}
Live Avatar~\cite{huang2025liveavatar},
SoulX-FlashTalk~\cite{shen2025flashtalk},
SoulX-LiveAct~\cite{zhen2026liveact},
TalkingMachines~\cite{low2025talkingmachines},
StreamAvatar~\cite{sun2026streamavatar} and
Avatar-Forever~\cite{li2026avatarforever} animate faces from speech with
5--22B models on one to eight GPUs, several with a dedicated decoder GPU. Humans
answer after a mean 208\ms~\cite{stivers2009turntaking}; from the end of user
speech, Full-Duplex-Bench~\cite{lin2025fullduplexbench} measures 0.27\,s for
Moshi~\cite{defossez2024moshi} and 1.30\,s for a commercial voice agent.

\paragraph{Diffusion serving.}
StreamDiffusionV2~\cite{feng2026streamdiffusionv2} serves streaming
video-to-video with a pipeline across GPUs and a host-managed ring-buffer cache, and
SlackServe~\cite{xia2026slackserve} manages resources and quality across
streaming video requests. xDiT, PipeFusion and
DistriFusion~\cite{fang2024xdit,fang2025pipefusion,li2024distrifusion}
parallelize large images and videos for offline requests, and
USP~\cite{fang2024usp} long-context sequences.

\paragraph{Overlapping communication with computation.}
Our tensor partitioning follows Megatron-LM~\cite{shoeybi2019megatron}.
Communication has been overlapped with computation by fusing collectives into
GEMM kernels~\cite{chang2024flux,zheng2025tilelink,zheng2025tritondistributed},
with hardware support~\cite{pati2024t3}, and by pipelining decomposed
operations~\cite{zhu2025nanoflow,wang2024asynctp}. ParallelKittens~\cite{sul2026parallelkittens}
fuses GEMM and all-reduce with in-switch reduction, and we extend its Hopper
kernel; NCCL offers NVLS collectives~\cite{nvidia2026nccl,nvidia2026nvls}; TensorRT-LLM~\cite{nvidia2026tensorrtllm}
ships a fused GEMM + all-reduce; concurrent work extends it to visual
generation with FP8 block scales~\cite{nvidia2026gemmallreduce} and builds
multi-node fused kernels~\cite{mao2026mkernel}. We contribute an FP8 variant for odd TP degrees
with a shared tail queue, a Blackwell port, and its behavior inside a
compiled, CUDA-graphed production model.

\paragraph{KV caches, positions and quantization.}
PagedAttention~\cite{kwon2023vllm} and attention
sinks~\cite{xiao2024streamingllm} shaped KV management for LLM serving.
Caching keys after RoPE~\cite{su2024roformer} and shifting them is established
in LLM engines~\cite{llamacpp2023kshift,huggingface2024cache}, as are FP8 and
low-bit KV caches~\cite{vllm2026fp8kv,liu2024kivi,hooper2024kvquant} and FP8
attention with per-head descales~\cite{shah2024flashattention3,dao2026fa3code}. Concurrent
work fuses QK-norm, RoPE and the cache write~\cite{flashinfer2026qknormrope}
and precomputes AdaLN tables~\cite{li2026solh3}. In our
model the positions slide every step and reset every 42 steps, which is why
the launch configuration re-applied RoPE at read time; caching post-RoPE keys
needed the delta rotation, and CUDA-graph capture a device-side write
position.

\section{Conclusion}
\label{sec:conclusion}

Serving a 57B video diffusion transformer to one user at a time, at a
sustained 166.7\ms{} per chunk, differs from serving it for throughput. At batch size one the cost that matters is work repeated in each
of a step's 240 attention calls over a cache that is mostly unchanged, and
the communication in series with each block. Moving that work to write time
took the diffusion step from 240.9 to 159.3\ms{} and halved the GPUs per
session; folding the all-reduce into the GEMM reached
147.4\ms, with no detected degradation in lip sync. In production the
worker averages above 23.5\,fps in 99.5\% of sessions of at least a
second, and of a 2.1\,s turn our pipeline takes half a second. What remains is
mostly outside the step: the voice agent, the pipeline depth that delays
barge-in, and a fleet sized for its peak.

\clearpage
\appendix
\section{Measurement methodology}
\label{app:repro}

All measurements come from four tools; their raw data, traces and output
videos are archived internally. Our model weights and production telemetry
are proprietary; Appendix~\ref{app:recipes} lists what is needed to rerun the
external systems.
\begin{itemize}
\item \textbf{Session benchmark.} Drives the production worker's components in
production order (setup, warm-up, per-session reset, AR step loop fed from
recorded speech) on the real topology, records per-step host and CUDA-event
timings on every rank, and optionally captures profiler traces of a step
window. Used for the live-length sessions, the waterfall and the Blackwell
runs; the fused kernel has a separate microbenchmark.
\item \textbf{Quality scorer.} Generates each evaluation clip with its
reference image, prompt and speech, then scores lip sync (SyncNet, Euclidean
distance, video resampled to 25\,fps, with our own crop instead of S3FD) per
arm, and PSNR / SSIM per pair of
arms overall and per 10\,s window over the frames that differ (identical
frames are counted separately), with bootstrap intervals over clips.
\item \textbf{Turn analyzer.} Joins the server's trace spans with the
browser's trace using the client's clock-synchronization offset, follows each
turn through its first speaking latent, re-anchors GPU spans to host events on
the same rank, excludes turns that overlap the profiler window, and times
barge-ins from the voice agent's interruption signal.
\item \textbf{Telemetry puller.} Reads every Prometheus counter and histogram
bucket per series at weekly boundaries and sums each series' last value minus
its value a week earlier (a new series counts from zero; no counter reset
occurred); fractions are at bucket edges. The metric populations are observed,
not a reconciled census: in one week sessions of at least a second outnumber
sessions with frames by 2\%.
\end{itemize}

\section{Playback simulation}
\label{app:slack}
The simulator replays a session's measured unpaced step periods through one
generation stage and a 24\,fps player; decoding and delivery add a constant
offset and are not modeled. With slack $s$, step $k$ starts generating once
step $k{-}1$ is done and step $k{-}s$ has started playing: $s{=}1$ overlaps
generating a step with playing the previous one, and $s{=}2$ also queues one
finished step (166.7\ms{} of video). Playback starts, with nothing queued,
when step 0 is ready; a step that misses its slot stalls playback and shifts
every later slot, as the pacer does. In the production configuration, with
$s{=}1$ a 161\ms{} step that starts when its predecessor starts playing is
ready 5.7\ms{} before its slot, so a 263\ms{} reset step is 96\ms{} late and
playback stalls (${\approx}$10 stalls per minute of ${\approx}$100\ms, some
from slow ordinary steps). With $s{=}2$ the generator gains 5.7\ms{} per step
until it leads its slot by 172\ms; a reset cuts the lead to 70\ms, which is
rebuilt within 18 steps, before the next reset, so there are no stalls.

\section{Quality cohorts and long sessions}
\label{app:quality}
A clip enters the lip-sync comparisons only if SyncNet returns a finite
score for it in every run. SyncNet finds no track in one photoreal clip in any
run, and tracks one non-human clip in five of the seven runs with LSE-D
12.1--14.2 (12.1--13.7 across production runs alone); counting that clip
wherever both runs of a pair score it moves the bf16-cache LSE-D difference to
+0.06 [$-$0.01, 0.16]. Table~\ref{tab:qualityabs} gives absolute means on these
51 clips.

\begin{table}[h]
\caption{Absolute lip-sync means over the 51 clips SyncNet scores in every run.}
\label{tab:qualityabs}
\centering
\footnotesize
\setlength{\tabcolsep}{4pt}
\begin{tabular}{@{}lrr@{}}
\toprule
Run & LSE-D & LSE-C \\
\midrule
Baseline & 8.294 & 7.321 \\
Production & 8.270 & 7.353 \\
Production reruns & 8.261--8.285 & 7.345--7.362 \\
Fused & 8.293 & 7.336 \\
bf16 cache & 8.288 & 7.342 \\
\bottomrule
\end{tabular}
\end{table}

The long-session test alternates speaking and listening periods of 4--14 and
4--16\,s, near the 52--61\% listening of our live sessions. It has four
sessions per arm, one per character and calibration, each with one seed shared across arms. Lip sync is
scored per speaking period of at least 4\,s (SyncNet tracks 92--94 of the 130 in
both runs of a pair). The intervals in
\S\ref{sec:eval-quality} come from a bootstrap over these segments, which
treats segments of one session as independent; Table~\ref{tab:longsess} gives
the effects per session.

\begin{table}[h]
\caption{Long sessions: mean paired LSE-D difference from the first
production run per session, with the number of periods SyncNet scores in both
runs. Periods: speaking periods (scored by SyncNet in production).}
\label{tab:longsess}
\centering
\footnotesize
\setlength{\tabcolsep}{4pt}
\begin{tabular}{@{}llrrr@{}}
\toprule
Character & Calibrated & Periods & bf16 cache & Rerun \\
\midrule
Photoreal & listening & 33 (33) & $-$0.07 (33) & +0.01 (33) \\
Photoreal & speaking & 32 (32) & $-$0.01 (32) & $-$0.05 (32) \\
Non-human & listening & 33 (22) & +0.22 (15) & +0.17 (16) \\
Non-human & speaking & 32 (14) & +0.06 (12) & +0.23 (13) \\
\bottomrule
\end{tabular}
\end{table}

\section{Tensor-parallel shapes}
\label{app:shapes}
Column- and row-parallel weights are split on whole heads, without padding.
At TP-7 and TP-8 the split is uneven, so the ranks with more heads set the
step. FLOP and peak figures count only the model's own work, as there is no
head padding. The kernel microbenchmark uses one $K$ on every rank, the
average slice rounded up to the 128-element reduction block; at TP-7 it is
4.4\% below the larger ranks' $K$.

\section{External systems}
\label{app:recipes}
\emph{StreamDiffusionV2}: code at commit \texttt{6961a5cf}, weights
Wan2.1-T2V-14B \texttt{a064a6c7} and StreamDiffusionV2 \texttt{2373eb2b},
PyTorch 2.6, its four-GPU pipeline with \texttt{--schedule\_block}, 1 or 4
steps, upstream defaults otherwise; our patch adds per-chunk timestamps and a
clean exit, and keeps the input video on the host, copying each chunk to the
GPU when rank 0 picks it up, as a live stream would. Output chunks are matched
to inputs by message index, which we have not checked against the output
misalignment its decoder shows on four GPUs. \emph{Live Avatar}: code at commit \texttt{c3c47d03},
weights Live-Avatar \texttt{92cdccd1} and Wan2.2-S2V-14B \texttt{dab4e9c5},
PyTorch 2.8 with \texttt{torch.compile}, its \texttt{--fp8} option, Euler
sampler without guidance, 48-frame clips in 12-frame blocks; our patch adds
timestamps, optional pacing and a condition-cache size fix needed above
384$\times$704. For the paced latency of \S\ref{sec:eval-e2e} (384$\times$704,
five GPUs: four DiT ranks, one per denoising pass, and one decoder rank), rank 0
starts block $g$ no earlier than $t_0 + 0.48\,g$\,s, as if its audio had just
arrived whole, and latency runs to the block's frames on the host; audio
features are computed from the whole file before generation; the audio is the first
60\,s of our evaluation speech. The four-GPU
runs used the box of Table~\ref{tab:oss}, the five-GPU run an 8$\times$H100 node.

\section{Fused kernel details}
\label{app:kernel}

\begin{figure*}[t]
\centering
\includegraphics{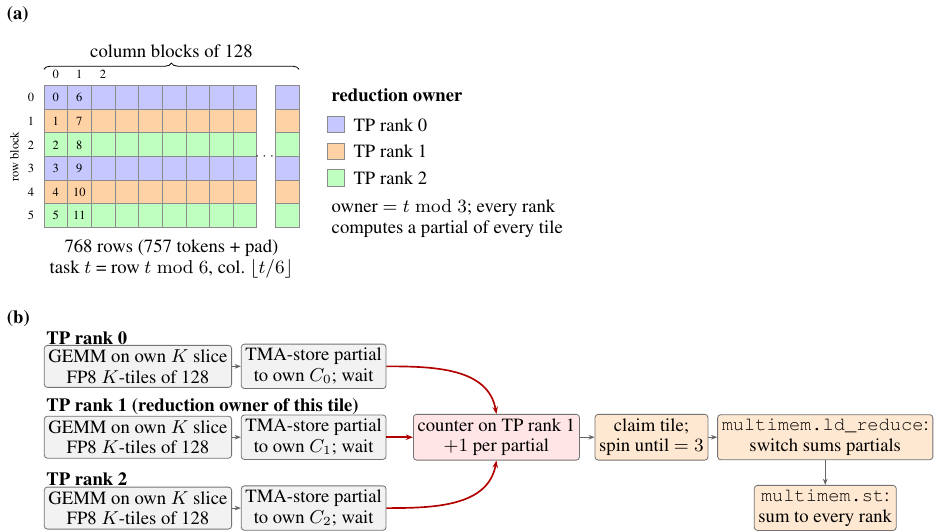}
\caption{The fused FP8 GEMM + NVLS all-reduce on Hopper at the production
shape (TP-3). (a) The 757-row output is covered by 128$\times$128 tiles in six
row blocks, with storage padded to 768 rows. Compute CTA $b \in [0, 129]$ (of
132 SMs) processes tasks $t = b + 130j$ in column-wise order. Every rank computes a partial for every tile;
$t \bmod 3$ assigns the reduction owner. (b) For one tile, every rank multiplies its $K$ slice,
TMA-stores the bf16 partial into its own symmetric buffer, waits for the store
to complete and adds one to the tile's counter on the owner. The owner's
reduction warps (two communication CTAs, joined by each compute CTA once it
runs out of GEMM work) claim owned tiles in order from a work queue, wait until
all three partials have landed, and reduce the tile in the NVSwitch with
\texttt{multimem.ld\_reduce}, writing the sum to every rank with
\texttt{multimem.st}. Early tiles are reduced while later tiles are still in
the GEMM. After the call an epilogue kernel resets the counters and holds all
ranks at an exit barrier, so no rank overwrites a partial a peer has not yet
reduced.}
\label{fig:kernelproto}
\end{figure*}

\paragraph{Numerics.} The GEMM is bit-exact against \texttt{torch.\_scaled\_mm}
with the same scales; the reduced output differs from GEMM + NCCL by bf16
rounding of the sum (at most ${\approx}$1 bf16 ulp of the output scale, mean
absolute difference 0.005 at output magnitude 10).

\paragraph{Hopper tuning.} Five pipeline stages instead of four
(163.9 $\rightarrow$ 154.9\us): the GEMM waits on weight tiles, not tensor
cores. The activation-scale vector is padded to whole 128-row tiles, since each
consumer warpgroup of the last row-block loads its full slice; a 64-row pad
read past the tensor and faulted only at $M{=}1671$. ThunderKittens must be
built with \texttt{-fno-strict-aliasing}: it builds TMA descriptors in host
code by type punning, and strict aliasing at \texttt{-O3} produced descriptors
that fault inside the kernel, with byte-identical device code.

\paragraph{Blackwell deadlock and tracing.} The deadlock appeared as a silent
hang at 100\% GPU utilization. ThunderKittens' tensor-memory allocator waits
on a cluster barrier in its destructor, so the post-GEMM reduction must run
inside the allocator's scope, or warp 0 (in the destructor) and the other
warps (at the reduction's barrier) wait on each other. Separately, the
compiled path must not import the extension module, which the tracer cannot
follow under \texttt{fullgraph=True}; the eager preparation step publishes the
kernel geometry for the traced path to read.

\section{Cache update and position reset}
\label{app:reset}
Each of the 240 caches (60 layers $\times$ 4 denoising passes) holds 1,246
global slots, written once per session, and a ring of six latent slots of 757 tokens: 5,788 slots, all read by
attention. In every step each block computes Q,
K and V for the 757 new tokens, RMS-normalizes Q and K, applies RoPE at the
new latent's temporal index, quantizes K and V with the cache's per-(layer,
head) scales and writes them into the slot at the device-side write position,
then attends over the whole cache. Scales are set once per session, on entering
the seventh step, from the amax of the bf16 cache the first six filled, times 1.5.

Every 42 steps every cached frame's temporal index moves by $\Delta = -42$.
A window frame at temporal index $t$ has angle $\theta_j(t) = \omega_j (a t + b)$
per RoPE frequency $j$, and the global tokens' positions do not depend on the
window. At a
reset, before each denoising pass's blocks read their caches, the window keys are
dequantized with their head's scale, rotated by $\omega_j a \Delta$ in fp32
(spatial components are unchanged), clamped to the FP8 range and requantized
with the same scale; values are not touched. The window (6 latents) is shorter
than the reset period (42 steps), so no key is requantized twice.

On B200, cuDNN's MXFP8 attention takes per-32-element block scales instead of
per-head ones, but blocks values along the sequence, so each 757-token write
would share blocks with older tokens, and padding writes to whole blocks needs
a key mask that the wrapper and cuDNN build we use do not offer.

\section{Pixel divergence}
\label{app:pixels}
\begin{figure}[t]
\centering
\includegraphics[width=\columnwidth]{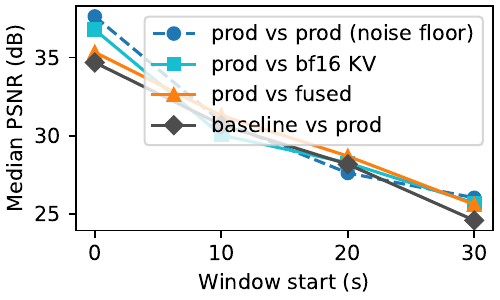}
\caption{Median PSNR between two runs per 10\,s window over the frames that
differ, over the clips long
enough for each window.}
\label{fig:quality}
\end{figure}

Two production runs, and production versus a bf16 cache, are bit-identical
for the first 21 frames (0.875\,s) of every clip and reach 37.6 and 36.8\,dB
over the first 10\,s, against 34.7--35.3\,dB for configurations that change
numerics from the first step; after 10\,s every pair sits near 30, then 28,
then 25\,dB. PSNR means leave identical
frames out, whole-clip and per window. Whole-clip PSNR between configurations ranges from 32.6 to
35.5\,dB, with two production runs the most similar. Per 10\,s window
(Figure~\ref{fig:quality}) the difference is only \emph{when} runs diverge.

\section{Published end-to-end latency}
\label{app:e2e}
\paragraph{Harness.} \label{app:harness} The worker (LiveKit Agents 1.8.3 and
the avatar's official plugin) plays 30 clips cut at pauses from our evaluation
speech, 10 each of about 3, 8 and 15\,s, in five sessions of at most 120\,s,
each opened by a warm-up turn that is excluded and started 5\,s after the
avatar's video appears. It sends each clip as 20\ms{} frames of 16\,kHz audio
as fast as the plugin accepts them, as streaming speech synthesis would, with
transcript synchronization off. A viewer in the same LiveKit Cloud room, on the
same host and clock, marks the avatar's voice at the first two consecutive
10\ms{} frames above $-$40\,dBFS; latency is that time minus the agent's
first sent frame and the clip's own lead-in, found with the same detector. A
mock avatar that returns the audio after 300\ms{} measured 412\ms{} through
the cloud (112\ms{} of overhead), and 367 and 882\ms{} for 300 and 800\ms{}
on a local server.

\paragraph{Published figures.} Table~\ref{tab:e2e} lists the published
latency figures, each over the interval its source defines, next to ours over
the closest interval we measure. The intervals seldom match exactly, so the
table is context rather than a comparison.

\begin{table*}[t]
\caption{Latency of real-time avatar systems as each source defines it, and
ours over the closest measured interval. n/r: not reported.}
\label{tab:e2e}
\centering
\footnotesize
\setlength{\tabcolsep}{3pt}
\renewcommand{\arraystretch}{1.05}
\begin{tabular}{@{}>{\raggedright\arraybackslash}p{4.1cm}>{\raggedright\arraybackslash}p{4.9cm}r>{\raggedright\arraybackslash}p{2.0cm}>{\raggedright\arraybackslash}p{3.0cm}@{}}
\toprule
System & Interval (as defined by the source) & Latency & Voice and language & Setting \\
\midrule
\multicolumn{5}{@{}l}{\emph{Video side: speech audio in $\rightarrow$ first video frame out}} \\
\textbf{Runway Characters}, benchmark & audio chunk $\rightarrow$ its frames on host & 293\ms{} p50 & -- & 4$\times$H100, 704$\times$1088, 24\,fps \\
\textbf{Runway Characters}, live & agent's first audio $\rightarrow$ first speaking frame sent & 493\ms{} p50 & -- & same \\
\textbf{Runway Characters}, harness (\S\ref{sec:eval-e2e}) & agent's first audio frame $\rightarrow$ avatar's voice at a viewer & 557\ms{} p50 & -- & public API, LiveKit Cloud \\
AVTR-1~\cite{kravtsov2026avtr1} & agent's first speech $\rightarrow$ first speaking frame sent & 390--454\ms{} mean & -- & 1 GPU, 153M model, 25\,fps \\
VASA-1~\cite{xu2024vasa1} & ``preceding latency'' & 170\ms & -- & 1$\times$RTX 4090, 512$^2$, 40\,fps \\
SoulX-FlashTalk~\cite{shen2025flashtalk} & start-up, one generation cycle & 0.87\,s & -- & 8$\times$H800, 720$\times$416, 32\,fps \\
SoulX-LiveAct~\cite{zhen2026liveact} & ``latency'' (not defined) & 0.94\,s & -- & 2$\times$H100/H200, 20\,fps \\
StreamAvatar~\cite{sun2026streamavatar} & input chunk $\rightarrow$ first frame (incl.\ 0.48\,s buffer) & 1.20\,s & -- & 2$\times$H800, 928$\times$704, 25\,fps \\
Live Avatar~\cite{huang2025liveavatar} & audio arrival $\rightarrow$ first frame & 1.21\,s & -- & 5$\times$H100, 720$\times$400, 45\,fps \\
\midrule
\multicolumn{5}{@{}l}{\emph{Reported interactive-system response latency (endpoints differ)}} \\
\textbf{Runway Characters}, live & end of user speech $\rightarrow$ avatar audio at the client & 2.13\,s p50 & third-party agent & p10--p90 1.45--3.00\,s \\
LiveTalk~\cite{chern2025livetalk} & latency per turn (not defined) & 1.16\,s & Qwen3-Omni & video model 1 GPU (system n/r), 512$^2$, 25\,fps \\
Wan-Streamer v0.2~\cite{huang2026wanstreamer2} & user unit $\rightarrow$ response, + 350\ms{} assumed network & ${\approx}$0.55\,s & integrated model & 640$\times$368, 25\,fps \\
Commercial voice agent~\cite{lin2025fullduplexbench} & end of speech $\rightarrow$ start of response & 1.30\,s mean & -- & Full-Duplex-Bench \\
\bottomrule
\end{tabular}
\end{table*}

\end{document}